\documentclass[12pt]{amsart}
\usepackage[utf8]{inputenc}
\usepackage[a4paper,margin=1in,footskip=0.25in]{geometry}

\usepackage{float}
\usepackage{graphicx}
\usepackage{amssymb}
\usepackage{epstopdf}
\usepackage{amsmath}
\newtheorem{thm}{Theorem}
\usepackage{subfig}
\usepackage[authoryear]{natbib}
\usepackage[ruled,vlined,linesnumbered]{algorithm2e}
\usepackage{comment}
\usepackage{booktabs}

\usepackage[colorlinks=true,linkcolor=blue, allcolors=blue]{hyperref}%

\usepackage{algpseudocode}

\begin{document}
\author{Shane Lubold}
\author{Bolun Liu}
\author{Tyler H. McCormick}
\thanks{Contact information: Shane Lubold (sl223@uw.edu), Bolun Liu (bolun599@uw.edu), Tyler H. McCormick (tylermc@uw.edu). Research reported in this publication was supported by the National Institute Of Mental Health of the National Institutes of Health under Award Number DP2MH122405. The content is solely the responsibility of the authors and does not necessarily represent the official views of the National Institutes of Health.}

\address[Shane Lubold]{Department of Statistics, University of Washington}
\email{sl223@uw.edu}
\address[Bolun Liu]{Department of Statistics, University of Washington}
\address[Tyler H. McCormick]{Department of Statistics and Sociology, University of Washington}

\title{Spectral goodness-of-fit tests for complete and partial network data}

\begin{abstract}
Networks describe the, often complex, relationships between individual actors.  In this work, we address the question of how to determine whether a parametric model, such as a stochastic block model or latent space model, fits a dataset well and will extrapolate to similar data. We use recent results in random matrix theory to derive a general goodness-of-fit test for dyadic data. We show that our method, when applied to a specific model of interest, provides an straightforward, computationally fast way of selecting parameters in a number of commonly used network models. For example, we show how to select the dimension of the latent space in latent space models. Unlike other network goodness-of-fit methods, our general approach does not require simulating from a candidate parametric model, which can be cumbersome with large graphs, and eliminates the need to choose a particular set of statistics on the graph for comparison.  It also allows us to perform goodness-of-fit tests on partial network data, such as Aggregated Relational Data. We show with simulations that our method performs well in many situations of interest. We analyze several empirically relevant networks and show that our method leads to improved community detection algorithms. R code to implement our method is available at \url{https://github.com/slubold/Network_GOF}. 
\end{abstract}

\maketitle

\section{Introduction}
Networks consist of connections, also known as edges or ties, between individual actors or nodes.  Such data are common in a variety of settings in the social, economic, and health sciences.  Informal insurance \citep{ambrusms2012, cai2017interfirm}, education decisions \citep{calvo2009peer}, sexual health \citep{handcock2004likelihood}, international trade \citep{chaney2014network}, and politics \citep{diprete2011segregation} are among the many settings in which  networks play a major role. 
The past decades have seen a flurry of parametric, statistical models to characterize network structure.  The simplest model is the Erdös-Rényi model \citep{ER}, in which edges form independently with the same probability. More complex models have been developed, such as stochastic block models (SBM) and degree-corrected variants (see, among many others, \citet{Holland_SBM}, \citet{airoldi2006mixed}, \citet{ rohe2011spectral}, and \citet{yan2014model}), latent space models \citep{hoff2002latent, Hoff_Bilinear, shalizi2017consistency, Lubold2020}, exponential random graph models (ERGMs) \citep{holland1981exponential, Hunter_GOF}, and many more.  Among others, \cite{Feinberg_survey} provides a survey of common network models. 

Given the multiple available models, a natural question in practice is how to choose a model that is appropriate for a particular dataset.  Broadly speaking, there are two common approaches to this problem currently in the literature.  A first approach leverages the fact that the problem has a ``parametric null.'' Since many network models are also generative, one common strategy is to estimate parameters of the model in question, then simulate a series of graphs.  
Statistics from the fitted model should resemble the observed statistics \citep{Ouadah, Onnela, Shore2015, Gao_Witten}. A potential issue with these methods is that, in many setting, there is limited information available to a practitioner to decide which statistics to use for comparison.  In some cases, the researcher can choose statistics that are important to their application, but it might not always be possible to select \emph{a-priori} which statistics will be the most important in future analyses.  This method also requires taking multiple samples from the generative process, which can be cumbersome in high dimensional settings.

A second common strategy for assessing goodness-of-fit (GoF) involves using the penalized likelihood methods, such as the Bayesian Information Criterion (BIC) or Akaike Information Criterion (AIC).  For example, AIC or BIC could be used to select the dimension of the latent space in latent space models, but as we show with simulations in Appendix \ref{sec: LS_Appendix}, the BIC approach leads to poor dimension estimates. In fact, the manual for one of the most common software packages for fitting parametric models, \texttt{latentnet}, states ``\textit{It is not clear whether it is appropriate to use this BIC to select the dimension of latent space ...}."  This issue has also been documented previously \citep{oh2001bayesian, Tantrum, Raftery07, lenk2009simulation, gormley2010mixture}. By contrast, the goodness-of-fit test we propose here does not use a penalized likelihood to select dimension. Instead, it uses the eigenvalues of a random matrix that measures how well an assumed model fits the data. This procedure, as we show in this work, outperforms BIC and similar metrics when applied to selecting the dimension of the latent space.

In this work, we present a novel goodness-of-fit test to assess model fit when the network of interest is un-directed or directed.  Our method also accommodates partial network data, which is a vital part of modern network analysis \citep{bernard2010counting, mccormick2010many, breza2017using, Breza_consist, Leung}, but are not easily handed in existing goodness-of-fit tools.  Our goal is to derive a testing framework for the hypotheses
\begin{equation}
\label{eq: origina_H}
    H_0: G \sim F_\theta, \ \ H_a: H_0 \text{ is false} \;, 
\end{equation}
where $G$ is a random network of interest and $F_\theta$ is a parametric network model with an (unknown) parameter vector represented by $\theta$. In words, we have an assumed parametric network model, $F_\theta$, and our goal is to test whether $G$ could be drawn from $F_\theta$. Throughout this work, we will use graph and network interchangeably. 

A critical aspect in the setup of the tests above is that they assess goodness-of-fit for the entire model simultaneously.  In many settings, this means that the contribution of individual parameters to the fitness measure may not be separately identified.  Take, as an example, the latent distance model discussed above.  This model has both individual effects for each respondent and latent distances for each pair.  A common question, as described above, involves testing for the dimension of the latent space.  To test exclusively for the dimension, we would need to either marginalize over or condition on potential values for the additional model parameters (see the discussion in~\citet{oh2001bayesian}, for example, which makes this point in the related setting of multidimensional scaling).  In the latent space model this is particularly challenging since both the latent distances and individual effects impact overall graph properties, such as the density (see, for example,~\cite{Lubold2020} for further discussion).  In our approach, we ask a related, but distinct question from the literature that tests for specific model parameters.  In the case of the latent space model, for example, our test asks whether a model, overall, could have plausibly generated a given set of data, rather than attempting to identify a single ``true'' latent dimension.  Despite this, in our simulations, we see however that this approach tends to find the true dimension with high probability.

The motivation for our test statistic is taken from a result that has been used before in community detection \citep{Jing, Sarkar_Bickel} and two-sample tests for networks \cite{Chen2020}. The result, which goes back to \cite{Yau2012}, states that if $A$ is a $n \times n$ random symmetric matrix with (i) $E(A_{ij}) = 0$ and (ii) $\sum_{j \neq i} \text{Var}(A_{ij}) = 1$ for each $i$, then $n^{2/3}(\lambda_{\max}(A) - 2)$ converges in distribution to a random variable with a Tracy-Widom distribution, where $\lambda_{\max}(A)$ is the largest eigenvalue of the matrix $A$. The same  argument shows that the smallest eigenvalue of $A$, which we denote by $\lambda_{\min}(A)$, satisfies a similar central limit theorem.

Leveraging this result, we propose a two-step procedure to test the hypothesis in (\ref{eq: origina_H}). First, we compute an estimate $\hat \theta$ of $\theta$ and estimate $\hat P_{ij} := P(G_{ij} = 1 | \hat \theta)$, where $G_{ij}=1$ if person $i$ and person $j$ are observed to be connected in the network, and $P_{ij}$ is the probability of such a connection (as defined by the assumed parametric model). Second, we define the random matrix $A$ by, for $i \neq j$,
\begin{equation}
\label{eq: original_A}
    A_{ij} = \frac{G_{ij} - \hat P_{ij}}{\sqrt{(n-1)\hat P_{ij}(1-\hat P_{ij})}} \;,
\end{equation}
and $A_{ii} = 0$. Under $H_0$, we expect that a reasonable estimator for $\hat P_{ij}$ should approximate $P(G_{ij} = 1 | \theta)$, so $A$ from (\ref{eq: original_A}) should approximately satisfy conditions (i) and (ii). We can then compare the largest and smallest eigenvalues of $\hat A$ against quantiles of the Tracy-Widom distribution to construct a test of $H_0$ in (\ref{eq: origina_H}). 

Our paper contributes to the literature on testing goodness-of-fit for network models in three ways. 
First, we expand work by \cite{Jing}, which estimates the number of communities in a stochastic block model, to accommodate a variety of common parametric network models. Second, we develop a test for directed data by introducing a similar central limit theorem for eigenvalues of non-symmetric matrices from \cite{Johnstone} and \cite{Chafai}. Third, we show how to test (\ref{eq: origina_H}) when the researcher only has access to partial network data, such as Aggregated Relational Data.  Along with asymptotic arguments we also present a bootstrap procedure which improves performance of our hypothesis tests in finite samples.

The paper is structured as follows. First, we review relevant literature in the remainder of this section.  Next, we introduce the construction of the Tracy-Widom distribution and asymptotic arguments for un-directed, directed, and partial network data in Section~\ref{sec: methods}. In Section \ref{bootstrap}, we discuss a bootstrap correction algorithm to improve finite sample properties. We then present a series of network models that are compatible with our method in Section \ref{sec: model} along with simulation results. Lastly, in Section \ref{sec: real}, we analyze several observed networks using a latent distance model~\citep{hoff2002latent, Hoff2003, Hoff_Bilinear}, which assumes that relationships in the network depend on the positions of actors in latent ``social space'' of low but unknown dimension.  Our goal with these data is to test for the minimal latent dimension. The R code for the simulations and to implement the method can be found in \url{https://github.com/slubold/Network_GOF}.

\subsection{Literature Review}
Goodness-of-fit methods for dyadic data generally address the question ``Does the proposed model fit my network data well?" One reason this problem is challenging, among others, is that there is often only one network of interest. In other words, we cannot access more draws from the distribution that generated the observed network. Many goodness-of-fit methods for dyadic data try to use Monte Carlo methods to simulate network statistics, such as average degree or average path length. If the simulated values match the observed values, then one might claim that the fitted model is adequate. See, for example, \cite{Ouadah} which derives a test for an Erdös-Rényi network using the degree distribution.  \cite{Chao} looks at using small graph statistics, such as the number of triangles or edges, to determine if there is community structure in a network. Each of these methods, generally speaking, requires a new derivation of a central limit theorem for the network statistic of interest under a suitable null hypothesis, which makes a general method hard to derive.

\cite{Shore2015} proposes a general goodness-of-fit method based on resampling the graph Laplacian's eigenvalues and constructing confidence intervals based on these values. Their null model is always the Erdös-Rényi model, which may not reflect the complexity of observed data.  Similarly, \cite{Levina} proposed a way to do cross validation with network data. In terms of more specific methods, For example, \cite{Gesine} derives a test of the form of an ERGM (defined formally in Section \ref{sec: model}) using kernel stein discrepancy. To the best of our knowledge, network goodness-of-fit tests using partial data are not well-studied. Recent work on modeling partial network data, such as \cite{bernard2010counting}, \cite{breza2017using}, \cite{ Breza_consist}, \cite{Leung}, and \cite{ Chatterjee_Thresholding}, consider model adequacy using out of sample prediction, which may be appropriate in some circumstances but asks a fundamentally different question than goodness-of-fit. 

Finally, our methods draw on results from random matrix theory and its applications. See, among others, \cite{Tao} and references therein, for an introduction. Our method builds on the method presented in \cite{Jing} and \cite{Sarkar_Bickel}, which also use spectral properties to estimate the number of communities in a stochastic block model. See \cite{Yau2012} and \cite{Komlos} and their references for recent work on related central limit theorems for eigenvalues. \cite{Chen2020} proposes a two-sample test for network using the Tracy-Widom distribution, which is the same distribution that motivates our goodness-of-fit test statistics.

\section{Methodology}
\label{sec: methods}

We now outline the goodness-of-fit problem. Let $Y$ be an $n \times n$ matrix containing relationships between actors or nodes, which we label from 1 to $n$. In this work, we usually only consider $Y$ to be binary-valued, so   $Y$ represents a network on $n$ nodes.  We suppose that $Y$ is drawn from some distribution $F$.  The network goodness-of-fit question then asks whether a given set of observed data $Y$ could plausibly have been generated by $F$. In many cases, it is possible to index $F$ by some parameter $\theta$. We are therefore interested in testing the GoF hypothesis in (\ref{eq: origina_H}). When $F = F_\theta$, we write $P_{ij} = P(Y_{ij} = 1 | \theta)$ to mean the probability that nodes $i$ and $j$ connect, given the parameter $\theta.$

The methodology we present in this work to test (\ref{eq: origina_H}) requires an estimate of $\theta$. We can estimate $\theta$ via maximum likelihood estimation (MLE), for example.
Assuming we have estimated $\theta$ with $\hat \theta$, we can use the parametric form of $F_\theta$ to obtain a fitted distribution $F_{\hat \theta}$. From this distribution, we can estimate $\hat P_{ij} = P(Y_{ij} = 1 | \hat \theta)$, which is the probability that nodes $i$ and $j$ connect, given the parameter $\hat \theta.$  In the next section, we derive a testing framework to test the hypothesis in (\ref{eq: origina_H}). We discuss the cases of undirected, directed, and partially-observed networks in their own sections, since each case requires a different approach.

\subsection{Undirected Networks}
We first consider the case where $Y$ corresponds to an undirected binary matrix. The following result, which motivates our test statistic for (\ref{eq: origina_H}), states that the eigenvalues of a transformation of the adjacency matrix satisfy a central limit theorem. See \cite{Yau2012, Lee, Komlos, Jing, Wigner} for related results and discussion. Formally, this result combines results from \cite{Yau2012} and \cite{Lee} and is formulated in Lemma A.1 of \cite{Jing}, among other works.

\begin{thm} [Lemma A.1 of \cite{Jing}]
\label{thm: Undirected_CLT}
Let $Y$ be the adjacency matrix of a random graph on $n$ nodes with edges drawn independently with probability $P_{ij}.$ Define the $n \times n$ random matrix $A$ with entries
\begin{equation*}
    A_{ij} := \frac{Y_{ij} - P_{ij}}{\sqrt{(n - 1)P_{ij}(1-P_{ij})}}, \ \ A_{ii} = 0 \;.
\end{equation*}
Then, as $n \rightarrow \infty$,
\begin{align}
    t_1 := n^{2/3} \left(\lambda_{\max}(A) - 2\right) \overset{d}{\rightarrow} TW_1, \label{eq: t_1} \\
    t_2 := n^{2/3} \left(-\lambda_{\min}(A) - 2\right) \overset{d}{\rightarrow} TW_1 \label{eq: t_2}
 \end{align}
 where $TW_1$ is the Tracy-Widom distribution with parameter 1. 
\end{thm}

\begin{figure}[btp!]
    \centering
    \subfloat[\centering ]{{\includegraphics[width=7cm]{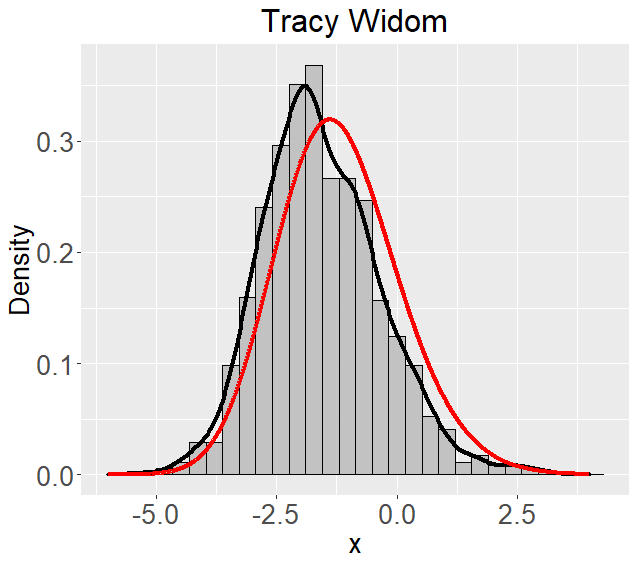} }}%
    \qquad
    \subfloat[\centering]{{\includegraphics[width=7cm]{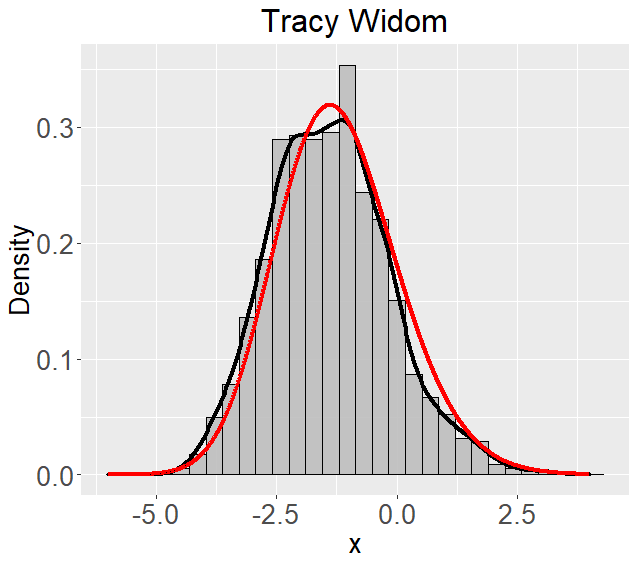} }}%
    \caption{\footnotesize{Distribution of statistic in Theorem \ref{thm: Undirected_CLT} for $n = 50$ (left) and $n = 1000$ (right), where the red curve corresponds to
the Tracy-Widom distribution with $\beta = 1$. The difference in the distributions decreases as $n$ increases, but the convergence is slow. This motivates the bootstrapping correction algorithm, given in Algorithm \ref{alg: boot_undirected}.}}%
    \label{fig:example}%
\end{figure}

In words, this result states the largest and smallest eigenvalues of the matrix $A$ satisfy a central limit theorem. In Figure \ref{fig:example} we plot the $TW_1$ distribution as well as $n^{2/3}(\lambda_{\max}(A) - 2)$ to illustrate this theorem.

In general when testing (\ref{eq: origina_H}), we do not know $\theta$, but as mentioned in the previous section, we do have an estimate $\hat \theta$. We therefore plug in $\hat P$ in place of $P$, where $\hat P_{ij} = P(G_{ij} = 1 | \hat \theta).$  We then define  \begin{equation*}
    \hat A_{ij} := \frac{Y_{ij} - \hat P_{ij}}{\sqrt{(n - 1)\hat P_{ij}(1- \hat P_{ij})}}, \ \ \hat{A}_{ii} = 0 \;.
\end{equation*}
This suggests a test of (\ref{eq: origina_H}) based on both $\lambda_{\max}(\hat A)$ and $\lambda_{\min}(\hat A)$, using the statistics $\hat t_1$ and $\hat t_2$, where the hat indicates that we replace the unknown $P_{ij}$ with the estimate $\hat P_{ij}.$

We reject $H_0$ in (\ref{eq: origina_H}) when 
\begin{equation}
\label{eq: test_t_hat}
    \max\{\hat t_1,\hat  t_2\} > TW_1(1 - \alpha/2) \ \  \text{or} \ \ \min\{\hat t_1, \hat t_2\} < TW_1(\alpha / 2) \;,
\end{equation}
where $t_1$ and $t_2$ are the test statistics from (\ref{eq: t_1}) and (\ref{eq: t_2}),  $TW_1(\alpha/2)$ and $TW_1(1 - \alpha/2)$  are the $(\alpha/2)\%$ and $(100-\alpha/2)\%$ quantile of the $TW_1$ distribution, respectively. If we instead use $t_1$ and $t_2$, this test has size $\alpha$ by a union bound argument. In practice, the test that uses $\hat t_1$ and $\hat t_2$ is size $\alpha$ if the eigenvalues of $\hat A$ converge quickly enough to the eigenvalues of $A$ in probability. The next result, which we do not believe has previously been reported in the literature, gives a rate at which this happens. 

\begin{thm}
\label{thm: diff}
If $n^{2/3}(\lambda_{\max}(\hat A) - \lambda_{\max}(A)) = o_P(1)$, then $n^{2/3}(\lambda_{\max}(\hat A) - 2) \overset{d}{\rightarrow} TW_1$. Furthermore, the test in (\ref{eq: test_t_hat}) has size $\alpha$ as $n \rightarrow \infty.$
\end{thm}

Theorem \ref{thm: diff} states that if we want to show that the test based on $\hat P$, rather than the unknown $P$, has size $\alpha$ as $n \rightarrow \infty$,  we must prove that $\lambda_{\max}(\hat A)$ converges fast enough to $\lambda_{\max}(A)$. This is problem specific and depends on the complexity of the graph distribution. \cite{Jing} shows under certain constraints, the conditions of Theorem \ref{thm: diff} hold in the case of a stochastic block model. To the best of our knowledge,  there is no work that verifies this condition in other, more complicated models, such as the latent space model. In the simulations in this work, we assume that this condition holds and see that in many models, we achieve an approximately size $\alpha$ test as $n \rightarrow \infty$. This suggests that in these models, the condition in Theorem \ref{thm: diff} holds, but we do not have a formal proof that Theorem \ref{thm: diff} holds in these models.

Before continuing, we comment on the term $P(G_{ij} = 1 | \hat \theta)$. In many models, such as the stochastic block model (SBM), this term is available in closed form in terms of $\hat \theta$. In other cases, such as exponential random graph models, this is not the case, since the graph model asserts a joint distribution over all pairs of edges.  We are not aware of a formula for marginal probability of a single edge in terms of the graph model. In these cases, we need to estimate the marginal probability matrix $P$. We present a simple method in Algorithm \ref{alg: boot_ERGM} to do this.

\subsection{Directed Networks}
In the case where $A$ is the adjacency matrix of a directed network, then Theorem \ref{thm: Undirected_CLT} will not be applicable, since 
the eigenvalues of $A$ are not guaranteed to be real. To test (\ref{eq: origina_H}) in the case of directed networks, we therefore introduce two central limit theorems for the singular values of non-symmetric random matrices, which always exist and are real. Both of these results assume a matrix with independent entries with (1) mean zero and (2) variance 1. Note that this differs slightly from the undirected case, where we required that the sum of the variance of entries in each row was 1. To satisfy conditions (1) and (2) in the directed case, we define the random matrix $\hat A$ with entries
\begin{equation}
\label{eq: hatA_directed}
    \hat{A}_{ij} := \frac{Y_{ij} - \hat{P}_{ij}}{\sqrt{\hat{P}_{ij}(1 - \hat{P}_{ij})}}, \ \ \ \hat{A}_{ii} = 0 \;.
\end{equation}
where again $\hat P_{ij} = P(G_{ij}= 1 | \hat \theta)$. Notice that there is no $(n-1)$ in the denominator of the expression for $\hat A.$ 


\begin{thm}[Theorem 1.1 of \cite{Johnstone}] 
Let $A$ be a $m \times n$ standard Gaussian random matrix such that
\begin{equation*}
    A_{ij} \sim_{iid} \mathcal{N}(0, 1) \ \ \text{ for all $1 \leq i \leq m$, $1 \leq j \leq n$}.
\end{equation*}
Let $s_{\max}(A)$ be the largest singular value of $A$. Then, if $m = m(n) \rightarrow \infty$, with $m \leq n$, and $\lim_{n \rightarrow \infty} m(n)/n = \gamma \in (0, 1]$, 
\begin{equation*}
    \frac{s_{\max}(A)^2 - \mu_{1, n}}{\sigma_{1, n}} \overset{d}{\rightarrow} \text{TW}_1 \;,
\end{equation*}
where $\mu_{1, n} = (\sqrt{n - 1}+\sqrt{m})^2$ and $\sigma_{1, n} = \sqrt{\mu_{1, n}}(1/\sqrt{n - 1} + 1/\sqrt{m})^{1/3}$. Moreover, if $\gamma > 1$,
then the result remains true up to the swap of the roles of m and n in the formulas.
\label{thm: asymm}
\end{thm}

Theorem \ref{thm: asymm} requires that the entries of $A$ be Gaussian, which is not the case when $A$ is the (binary) adjacency matrix of a random network. In Figure \ref{fig: asymm}, we show that the convergence claim in Theorem \ref{thm: asymm} still holds reasonably well when the entries of $A$ follow a Poisson binomial distribution. This suggests that we can use Theorem \ref{thm: asymm} to construct a test when the entries of $A$ are not Gaussian. 

For directed networks, Theorem \ref{thm: asymm} suggests that we take our test statistic to be $(s_{\max}(\hat A)^2 - \mu_{1, n}) / \sigma_{1, n}$ with $m = n$ and the rejection region to be $\{x: TW_1(\alpha / 2) < x < TW_1(1 - \alpha / 2)\}$, which is identical to the undirected case. Moreover, it is not necessary to restrict $m = n$, which indicates such a test statistic is also applicable on directed networks or networks for which we only have partial network data. We will elaborate on these ideas in a later section. 

Our second result states that the scaled singular of a non-symmetric random matrix, when suitably transformed as in (\ref{eq: hatA_directed}), converge to an exponential-type distribution. 


\begin{figure}[t]
\centering
    \includegraphics[scale = 0.5]{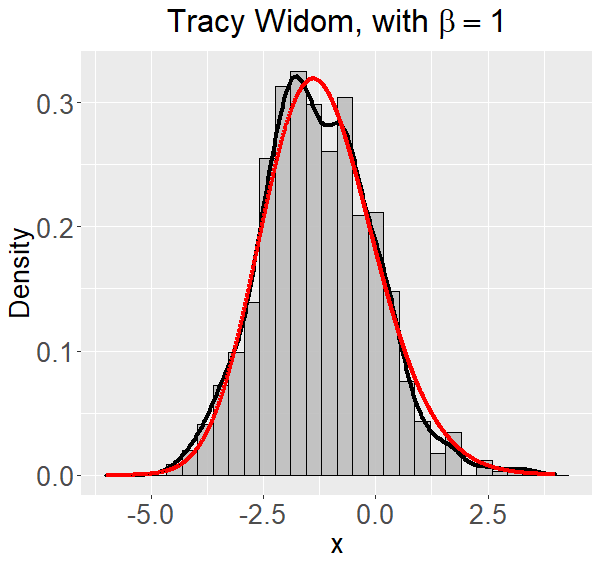}
    \caption{\footnotesize{ Distribution of the test statistics for networks with size $n = 1000$ in Theorem \ref{thm: asymm}. The red curve in the figure corresponds to the Tracy-Widom distribution with parameter $\beta = 1$. Overall, the convergence to the Tracy-Widom distribution is good enough and the theoretical Tracy-Widom distribution can be used for constructing our test statistic.}}%
\end{figure}

\begin{thm}[Theorem 2.4 of \cite{Chafai}]
\label{thm: exp_directed}
Suppose that $A$ is a random $n \times n$ non symmetric matrix whose entries have mean zero and variance 1. Then, if $s_{\min}(A)$ 
denotes the smallest singular value of $A$, 
\begin{equation*}
\lim _{n \rightarrow \infty} \mathbb{P}\left(\sqrt{n} s_{\min}(A) \geqslant t\right)=\exp \left(-\frac{1}{2} t^{2}-t\right).    
\end{equation*}
\end{thm}


Theorem \ref{thm: exp_directed} suggests that we take our test statistic to be $\sqrt{n} s_{\min}(A)$ and the rejection region to be $\{x : x > q_{\text{E}}(1 - \alpha)\}$ where $q_\text{E}(1-\alpha)$ is the $(1-\alpha) 100\%$ percentile of the  distribution in Theorem \ref{thm: exp_directed}.

To summarize, in this section we provided two central limit theorems for the singular values of random, non-symmetric matrices, which require that the entries of the random matrix have mean zero and variance 1. We discussed how to use the observed adjacency matrix $Y$ to construct such a matrix and to derive a test statistic for the GoF hypothesis in (\ref{eq: origina_H}). In Section \ref{sec: directed}, we discuss the performance of these two test statistics.

\subsection{Partial Network Data}
\label{sec: partial}

Suppose our goal, as it was above, is to test whether a graph $G$ is drawn from a particular model. That is, we want to test the hypothesis in (\ref{eq: origina_H}). In many applications,  complete network data is not available, is too expensive to collect, or cannot be collected for privacy-related reasons. A common form of partial network data, particularly in economics, is Aggregated Relational Data (ARD) \citep{breza2017using, Breza_consist, Leung}. In this work, we focus on using ARD to test the goodness-of-fit hypothesis in (\ref{eq: origina_H}), but we believe our framework can be extended to other data types too.

To describe what ARD is, suppose that we can partition the nodes of the network into $K$ categories $G_1, \dotsc, G_K$. These categories correspond to different covariates, so for example all nodes in $G_1$ have black hair and all nodes in $G_3$ are left-handed. We then ask $m \leq n$ nodes how many people they know with trait $j$ for $j = 1, \dotsc, K.$ In summary, by collecting ARD of a network with size $n$, we actually collect
$\{Y_{ij}: i = 1, \dotsc, m, j = 1, \dotsc, K\}$, with
\begin{equation}
    Y_{ij} = \sum_{k \in G_j} G_{ik}\;.
    \label{eq: ARD_def}
\end{equation}
We show in this section how to use ARD to test some of the network models previously mentioned. We assume, as is common in applications, that $|G_j|$ is known. Such information can come from census data or similar data sources.  

To illustrate ARD, we consider a simple example. Suppose that we consider $K = 2$ and $m = 4$ and we then collect the following data $Y$, written in matrix form as
\begin{equation}
\label{eq: Y_ARD}
    Y = 
    \begin{pmatrix}
    3 & 10  \\
    1 & 7 \\
    0 & 3 \\
    1 & 5
    \end{pmatrix} \;.
\end{equation}
This means, for example, that the first person we surveyed knows 3 people with trait 1 and the fourth person we surveyed knows 5 people with trait 2. In the above example, $m$ does not have to equal $K$ (and usually does not in practice), so $Y$ is often not square. This means that we cannot apply Theorem \ref{thm: Undirected_CLT} to test (\ref{eq: origina_H}). Instead, we test the hypothesis with Theorem \ref{thm: asymm}, which is applicable for non-square matrices. 

One challenge is to estimate $\theta$, given just the ARD. In this work, we consider a simple test of whether there is degree heterogeneity, which is equivalent to testing if the underlying model is an Erdös-Rényi model. Other, more complicated methods exist for estimating the parameters using only ARD in more complex models, such as those given in \cite{Leung} or \cite{breza2017using}.

\begin{figure}
    \centering
    \subfloat[\centering ]{{\includegraphics[width=7cm]{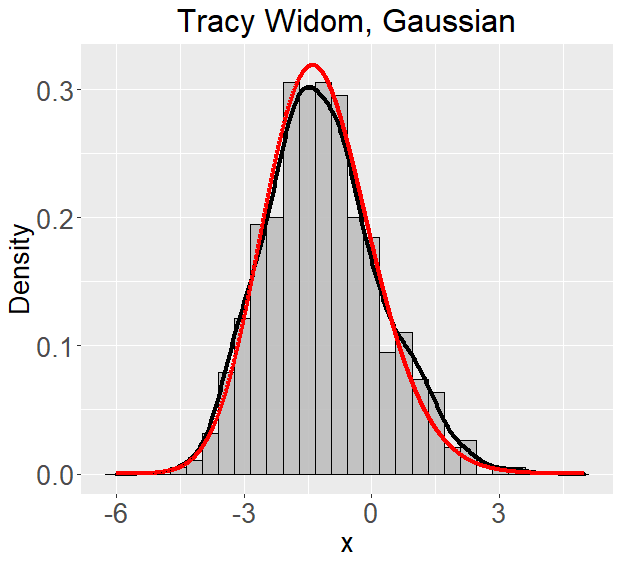} }}%
    \qquad
    \subfloat[\centering]{{\includegraphics[width=7cm]{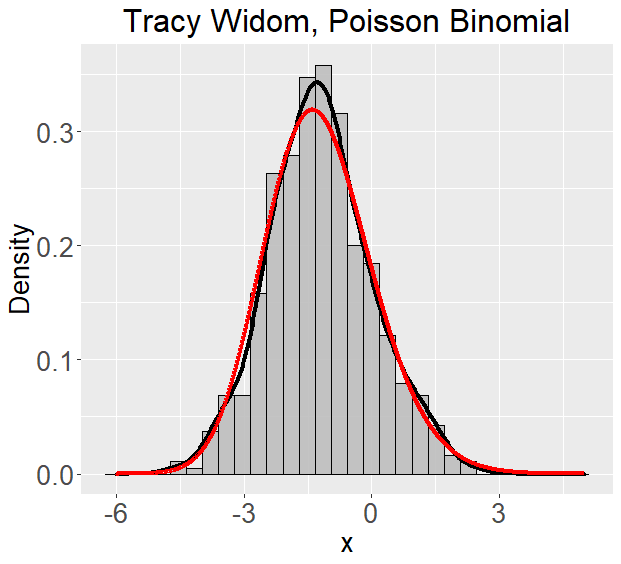} }}%
    \caption{\footnotesize{Left: Distribution of Tracy-Widom test statistics with parameter $\beta= 1$
 and $A$ to be a $600 \times 800$ standard Gaussian random matrix. Right: Distribution of Tracy-Widom test statistics with parameter $\beta= 1$ 
 and $A$ is a $600 \times 800$ random matrix re-centered with mean $0$ and variance $1$ from $G$ where $G_{ij}$ follows a Poisson binomial distribution with probability vector $p_i \sim_{i.i.d.} Unif(0, 0.3), i = 1, \cdots, 25$. The convergence of the distribution of Poisson binomial random matrix is 
almost as good as the Gaussian one, indicating Theorem \ref{thm: asymm} is compatible with non-Gaussian data, like Poisson binomial data.}}%
    \label{fig: asymm}%
\end{figure}

Before continuing, we discuss whether the assumptions in Theorem \ref{thm: asymm} hold for ARD.  We discuss three assumptions. First, recalling the notation from Theorem \ref{thm: asymm}, this result requires that $m \leq n$ so the matrix is ``long" rather than ``tall". In practice, the number of traits is smaller than the number of nodes we survey, so $Y$ is often ``tall", as it is in (\ref{eq: Y_ARD}). This does not pose a problem since the singular values for $Y$ and $Y^T$ are the same. Second, Theorem \ref{thm: asymm} requires that the $m / n \rightarrow \gamma \in (0, 1]$, which in the ARD context requires that the number of traits grows with $m$.  Previous work on the large sample properties of ARD estimators has either taken the number of traits as fixed (e.g.~\citet{Breza_consist}) or growing slowly, like $K = O(\sqrt{n})$ as in \cite{Leung}. Despite the assumption that $K$ grow with the sample size, our simulations in Section \ref{sec: ARD} show that this result of Theorem \ref{thm: asymm} still hold reasonably well. Lastly, in Theorem \ref{thm: asymm}, $A$ is required to be a Gaussian random matrix with continuous entries.  ARD are, however, counts.  If the number of ARD responses are relatively large, then the counts may appear reasonably normally distributed.  In most cases, however, we expect that the counts for most categories will be small.  Despite these potential violations of the required assumptions, Figure \ref{fig: asymm} shows that the approximation works well, at least visually, when the entries of the random matrix are not Gaussian and the underlying data come from a skewed distribution of counts, as would be the case in ARD.  We give simulation evidence in Section \ref{sec: ARD} that the approximation is sufficiently accurate to achieve favorable performance in hypothesis tests. 

\section{Bootstrap Correction}
\label{bootstrap}
As we saw previously, our asymptotic results can require a large sample size in practice.  We derive a bootstrapping correction method, similar to the one presented in \cite{Jing}. We note that our algorithm generalized the one in \cite{Jing}. Algorithm \ref{alg: boot_undirected} contains the algorithm for the undirected case, and Algorithm \ref{alg: boot_dir} contains the algorithm for the directed case. 

\begin{algorithm}

\SetAlgoLined
\KwIn{Observed sociomatrix: $G$; Estimated probability matrix $\hat{P}$; Bootstrap iterates: $B$; $TW_1$ mean: $\mu_{TW}$; $TW_1$ standard deviation: $s_{TW}$; Significance Level: $\alpha$.}

Compute 
    \begin{equation*}
        \hat A_{ij} = (G_{ij} - \hat P_{ij})/\sqrt{(n -1 )\hat P_{ij}(1 - \hat P_{ij})} \ ;
    \end{equation*}
    
\For{$b = 1$ \KwTo $B$}{

     Sample $G^\star_b \sim F_{\hat \theta}$\; 
    
     Compute
    \begin{equation*}
         [A^\star_b]_{ij} = ([G^\star_b]_{ij} - \hat P_{ij})/\sqrt{(n - 1)\hat P_{ij}(1 - \hat P_{ij})} \ ;
    \end{equation*}
    
    Set $\lambda_{b, \max}^\star = \lambda_{\max}( A^\star_b)$ and $\lambda_{b, \min}^\star = \lambda_{\min}( A^\star_b)$\;
    }
    
    Set $\mu_{\max}$ to be the sample mean of $\{\lambda_{b, \max}^\star\}_{b = 1}^B$ and $s_{\max}$ to be the sample standard deviation of $\{\lambda_{b, \max}^\star\}_{b = 1}^B$. Set $\mu_{\min}$ and $s_{\min}$ similarly\;
    
    Compute the test statistic $t$
\[t := \mu_{TW} + s_{TW} \cdot \max \left(\frac{\lambda_{\max}(\hat{A})-\mu_{\max}}{s_{\max}},-\frac{\lambda_{\min}(\hat{A})-\mu_{\min}}{s_{\min}}\right)\ ;\]

\uIf {$TW_1(\alpha / 2) < t < TW_1(1 - \alpha / 2)$} {
    Do not reject $t$ and set $\text{Rej} = \text{FALSE}$\;
} \Else {
    Reject $t$ and set $\text{Rej} = \text{TRUE}$.
}
\KwOut{Rejection of bootstrap statistic: Rej.}
 \caption{Bootstrap correction of Undirected Tracy Widom statistic}
 \label{alg: boot_undirected}
\end{algorithm}

Before continuing, we make a few remarks. First, the intuition behind this method is as follows: the distribution of $t' \equiv (\lambda_1(\hat A) - \mu_{\max})/s_{\max}$ is approximately $TW_1$ except that the mean and variance are incorrect, but by scaling $t'$ by $s_{TW}$ and then shifting $t'$ by $\mu_{TW}$ we obtain a better approximation of a $TW_1$ distribution. Second, we know that both $\lambda_{\max}(A)$ and $\lambda_{\min}(A)$ 
 have $TW_1$ distributions, and since we are taking a $\max$ over these two quantities, we want to use the $\alpha/2$ quantile of $TW_1$. This follows from a simple application of Bonferroni and leads to an $\alpha$-size test. Finally, in Appendix \ref{sec: Boot_Directed} we give a similar bootstrap correction algorithm for directed network data.

\section{Models}
\label{sec: model}
In this section, we demonstrate how our method can be used to perform model selection on a broad class of network models.  We consider the following problems:  
\begin{enumerate}
    \item Testing the link function in the $\beta$-model (Section \ref{sec: beta_model}).
    \item Comparing latent space models with different dimensions (Section \ref{sec: LS}).
    \item Comparing exponential random graph models with different forms (Section \ref{sec: ERGMS}).
    \item Testing degree heterogeneity using Aggregated Relational Data (Section \ref{sec: ARD}).
    \item Testing Community Structure in Directed Networks (Section \ref{sec: directed}).
\end{enumerate}

\subsection{Testing the Link Function in  \texorpdfstring{$\beta$}{Lg}-Model}
\label{sec: beta_model}

In this generative model, each node has a node effect $\beta_i$ which controls the probability it connects with other nodes. Let $\beta = (\beta_1, \dotsc, \beta_n) \in \mathbb{R}^n$ denote the vector of node effects. Then, conditioned on $\beta$, edges form independently in the undirected network with probability
\begin{equation}
\label{eq: beta_model}
    \mathbb{P}(G_{ij} = 1 | \beta) = \Lambda(\beta_i + \beta_j) \;,
\end{equation}
for some link function $\Lambda: \mathbb{R} \rightarrow [0, 1]$. Common examples of the link function include the expit and exp. \cite{diaconis} provides a fixed point method to compute the MLE of this model when the link function is the expit. We use this method to compute $\hat \beta$ in this work. The details of the MLE method can be found under Theorem $1.4$ of \cite{diaconis} and also in our supplementary R code. Note that many model selection tools, like BIC or AIC, would not be applicable here because the GoF question here is between two equally complex models because the only difference is in the link function. Our method therefore has the advantage of being applicable to link function tests. 

\begin{figure}
    \centering
    \subfloat[\centering ]{{\includegraphics[width=7cm]{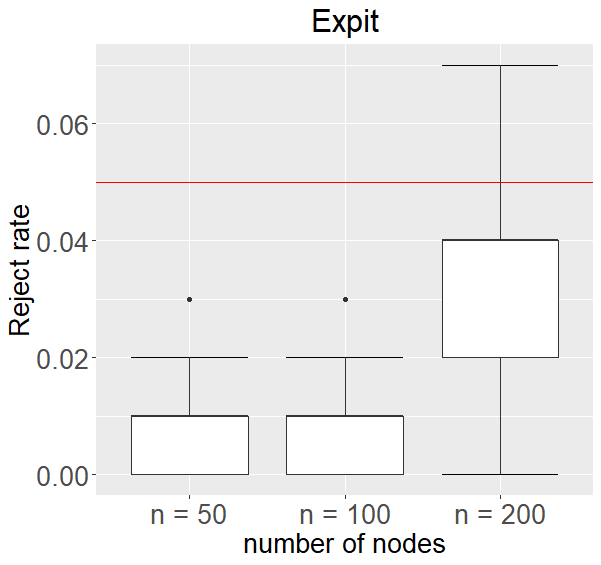} }}%
    \qquad
    \subfloat[\centering]{{\includegraphics[width=7cm]{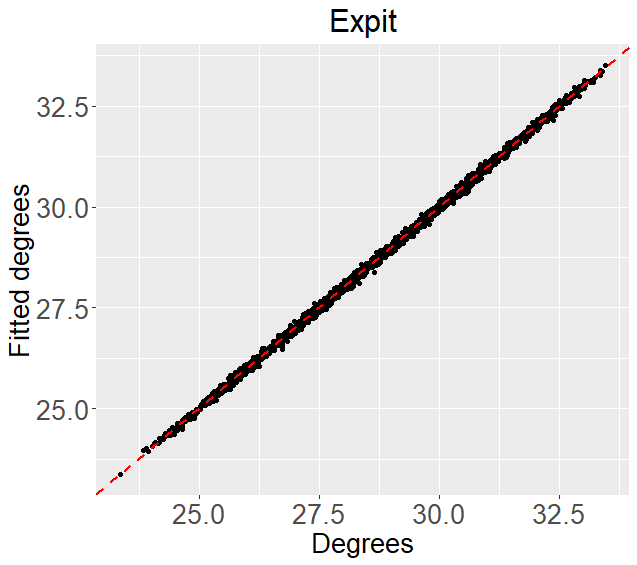} }}%
    \caption{\footnotesize{Left: Type I error for the null hypothesis in (\ref{eq: beta_null}) for $n = 50, 100, 200$. As $n $ increases, the Type I error increases to $\alpha = 0.05$. Right: On the $x$-axis we plot the average degree in networks of size $n = 200$, and on the $y$-axis we plot the average fitted degree across 50 simulations. We see that most points lie on the diagonal, which suggests that the expit model is a good fit. This is consistent with the left figure, which shows that our method is not rejecting (\ref{eq: beta_null}) often.}}%
    \label{fig: expit_data}%
\end{figure}

In our simulations, we consider three different cases. In the first case, we are interested in testing if $G$ is drawn from a $\beta$ model with the expit link function, or a Sigmoid function, such that $\text{expit}(x) = 1 / (1 + e^{-x})$. The hypothesis can be rewritten as
\begin{equation}
    H_0: G \sim \beta \text{-model with } \Lambda(x) = \text{expit}(x), \ \ \ H_{a}: H_0 \text{ is false} \;. 
    \label{eq: beta_null}
\end{equation}
To test this hypothesis, we generate a set $\beta$ of node specific effects on $n$ nodes and form a network with probabilities from (\ref{eq: beta_model}), with $\Lambda(x) = \text{expit}(x).$ We compute the MLE as described in \cite{diaconis} and form the $n \times n$ matrix $\hat P_{ij} = \text{expit}(\hat \beta_i + \hat \beta_j)$.
We then compute
\begin{equation*}
    \hat A_{i,j} = (G_{ij} - \hat P_{ij})/\sqrt{(n-1)\hat P_{ij}(1-\hat P_{ij})} \;.
\end{equation*}
for $i \neq j$ and $\hat A_{ii} = 0$. Using the bootstrapping algorithm from Section \ref{bootstrap}, we record the number of times that we reject $H_0$. We repeat this process 100 times and plot the type 1 \textcolor{red}{Type I} error for these 100 simulations in Figure \ref{fig: expit_data} for $n \in \{50, 100, 200\}$ with $\beta_i \overset{\text{i.i.d.}}{\sim} \text{Unif}(-2, 0).$

In our second set of simulations, we want to determine the power of our method for the hypothesis in (\ref{eq: beta_null}) when $H_0$ is false. In particular, we consider two reasons why $H_0$ is false. The first is that $\Lambda(x) = \exp(x)$, that is, the link function is incorrectly assumed. As before, we generate $\beta_i \overset{\text{i.i.d.}}{\sim} \text{Unif}(-2, 0)$ and generate the graph according to (\ref{eq: beta_model}) using $\Lambda(x) = \exp(x)$. In Figure \ref{fig: exp_data} we plot the rejection rates. We see that while the rejection rate is higher than in Figure \ref{fig: expit_data} (that is, when the null hypothesis is true), the average rejection rates for all $n$ values is below $10\%$. Therefore our method suggests that while the model is incorrectly specified, it is not a bad fit to the data.

In the third set of simulations, we generate $P$ with $P_{ij} \sim \text{Uniform}(0, 0.1)$ for $i < j$  drawn independently. We then generate an undirected network on $n$ nodes, where nodes $i$ and $j$ connect with probability $P_{ij}.$ In other words, there is no ``structure" to the matrix $P$, as there was when the matrix $P$ was formed according to (\ref{eq: beta_model}). In Figure \ref{fig: non_par_data}, we plot the rejection rates for (\ref{eq: beta_null}). We see that the power is much higher in this case than it was in the previous two simulations. In Figure \ref{fig: num_triangles_beta} we plot the number of triangles in the three fitted models versus the number of observed triangles. We see that the data drawn from the third simulation, where $P_{ij}$ are drawn uniformly, the simulated values do not match the observed values, but in the first two simulations we see a much closer fit.

\subsection{Testing Latent Space Models with Different Dimensions}
\label{sec: LS}

\begin{figure}[b]
    \centering
    \includegraphics[scale = 0.5]{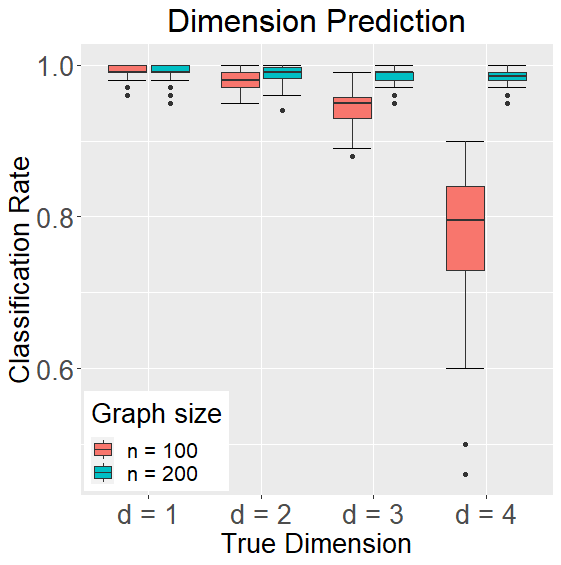}
    \caption{\footnotesize{Correct classification rate for $n = 100, 200$ for the dimension of the latent space in Section \ref{sec: LS}.  For a fixed $n$, increasing the dimension makes the problem harder and so the classification rate falls. However, the classification rate improves as we increase $n$ from 100 to 200.}}
    \label{fig: ls_dim}
\end{figure}

The latent space model, originally proposed in \cite{hoff2002latent}, is a generative network model that asserts that each node in a network has a position on some latent space. The closer two nodes in this latent space are, the more likely they connect. There is a large literature on latent space models. See, for example, \cite{Tantrum}, \cite{Hoff_Bilinear}, \cite{Dena}, \cite{oh2001bayesian}, \cite{shalizi2017consistency} and their references. In many cases, the user is interested in testing the dimension of the latent space as well as the geometry type. 
\cite{Lubold2020} shows how to, among other things, estimate the dimension of the latent space by using the clique structure in the network. In this work, we take a different approach and use the entire network to estimate the fit of the model to a hypothesized latent space dimension.

Let $G$ denote the adjacency matrix with observed covariate matrix $X$.  One form of the latent space model \citep{Hoff_Bilinear, Ma2020} asserts that conditioned on the network parameters, edges form independently with probability
\begin{align}
\label{eq: LS_model}
\begin{split}
    P(G_{ij} &= 1 | \theta) = P_{ij}, \ \ \ \ \ \text{where} \\
    \text{logodds} (P_{ij}) &=  \alpha_i + \alpha_j + \beta X_{ij} + \langle z_i, z_j\rangle,
\end{split}
\end{align}
where $\text{logodds}(x) = \log (x / 1 - x) $, $\{\alpha_i\}_{i = 1}^n$ are the parameters modeling degree heterogeneity, $\beta$ is the coefficient scaling the observed covariate $X$, $\langle z_i, z_j\rangle$ are the inner products between latent positions with $z_i \in \mathbb{R}^d$, and $d$ is the dimension of the latent space model. We let $\theta = (\alpha_1, \dotsc, \alpha_n, z_1, \dotsc, z_n, \beta)$ denote the collection of all the model's parameters.

Let $G$ be an observed network drawn from the model in (\ref{eq: LS_model}), where $z_i \in \mathbb{R}^{d_{\text{true}}}$. We develop a GoF procedure in Algorithm \ref{alg: dim_pred} via the Tracy-Widom statistic and a fast MLE method via non-convex projected gradient descent, described in \cite{Ma2020}.  The details can be found in Algorithm 1 and 3 in \cite{Ma2020} and also in the supplementary R code.  We now give a motivation for our algorithm. For any hypothesized dimension $d$, we can fit the model in (\ref{eq: LS_model}) and check if we reject the hypothesis that this dimension fits the network well. With no covariates or node effects, we expect to reject the null hypothesis for $d < d_{\text{true}}$, since a lower dimensional embedding should fail to accurately model the network structure, whereas dimensions equal to and higher than $d_{true}$ will capture the structure well and so we expect to fail to reject the corresponding hypotheses. 
This suggests that we should take the predicted dimension to be the smallest dimension for which we fail to reject the corresponding hypothesis.  As we mentioned in the introduction, the node effects have a confounding effect on the estimation procedure, so that model fits from two distinct dimensions, and their corresponding node effect estimates, might lead to equally good model fits. Even with the confounding issue, our simulations show that our procedure finds that the true dimension is often the smallest one that fits the model well.

 In the following simulations, we will focus on the inner product model without covariate components. However, our algorithm can be generalized to any inner product models with ``simple" covariates" as described in \cite{Ma2020} by following an almost identical methodology. For $n = \{100, 200\}$, we generated 50 sets of $\{\alpha, z\}$ with $d = \{1, 2, 3, 4\}$ respectively, where $\alpha_i \sim_{iid} \text{Unif}(-2, -1) \times 10^{-2}$ and $z_i \sim_{iid} \text{N}(0, I_d)$. For each combination of $\{n, d, \alpha, z\}$, 100 networks are drawn from the corresponding generated models and predicted with Algorithm \ref{alg: dim_pred}. The classification rates for each set of parameters are recorded and shown in Figure \ref{fig: ls_dim}. We notice two trends. First, as $n$ increases, the probability of correct dimension classification increases. Second, for a fixed $n$, larger dimensions are harder to classify correctly. This makes intuitive sense since higher dimensions often correspond to more complex latent space relationships,
and so it takes more data to model these relationships well.

\begin{algorithm}
\SetAlgoLined
  \KwIn{Observed sociomatrix: $G$.}
  Set $d_{\text{fit}} = 0$ and $T = 0$\;
  
  \While{$T = 0$}{
       Update $d_{\text{fit}}$ = $d_{\text{fit}} + 1$\;
       
       Compute the estimate $\hat{\theta}$ via Projected Gradient Descent algorithm with $d_{\text{fit}}$. Use $\hat \theta$ and the model in (\ref{eq: LS_model}) to compute $\hat P$\;
       
       Use $\hat P$ in the bootstrap algorithm (Algorithm \ref{alg: boot_undirected}) to determine if the null hypothesis $H_0: d_{\text{true}} = d_{\text{fit}}$ is rejected\; 
       
       \uIf{$H_0$ is rejected}{
       Set $T = 1$\;
       } \Else {
       Remain $T = 0$\;
       }
  }
\KwOut{Predicted latent dimension: $d_{\text{fit}}$.}
 \caption{Dimension prediction for Latent Space model}
 \label{alg: dim_pred}

\end{algorithm}

\subsection{Comparing exponential random graph models with different forms}
\label{sec: ERGMS}

Exponential random graph models (ERGMs) are a common choice to model complex network data. To perform inference on these models, one must estimate an often intractable normalizing constant, which makes inference challenging. Some authors have presented maximum pseudo-likelihood \cite{morris} and Monte Carlo estimation methods.  In this section, we show how to apply Theorem \ref{thm: Undirected_CLT} to test the form of an ERGM.

We now briefly review the form of ERGMs.
This model asserts that a random graph $G$ arises through the model
\begin{equation}
\label{eq: main_ERGM}
    P(G = g \mid \theta) = \frac{1}{c(\theta)}\exp\left(\sum_{i = 1}^K \theta_i h_i(g) \right)
\end{equation}
where $h_1, \dotsc, h_K$ are functions of the graph $g$ and $c(\theta)$ is the normalization constant. The user specifies the functions $h$ as well as the value $K$. Some examples of $h$ include $h(g) = \sum_{i < j} g_{ij}$, the number of edges in $g$, and 
\begin{equation*}
    h(g) = \sum_{i,j,k}^n g_{ij}g_{jk}g_{kj} \;,
\end{equation*} 
the number of triangles in $g$. Except in simple cases, the MLE for $\theta$, denoted by $\hat \theta$, is not available in closed form. We compute the MLE using the  \texttt{ERGM} package in \texttt{R}. 

Having estimated $\theta$, we now need to estimate the $n \times n$ matrix $P$, where $P_{ij} = P(G_{ij} = 1|\theta)$. In most models, there is a clear correspondence between $\theta$ and $P$. For example, in a latent space model without covariates, once we estimate $\theta = (z_1, \dotsc, z_n, \alpha_1, \cdots, \alpha_n, \beta)$, we can simply use the graph model in (\ref{eq: main_ERGM}) to estimate $P$. But for ERGMs, the model in (\ref{eq: main_ERGM}) asserts a model for the entire network $G$ all at once, rather than specifying individual edge probabilities. To simulate $P$ from $\hat \theta$, we therefore propose to simulate from the fitted model and record the number of edges between pairs of nodes across $B$ simulations. We present this simple procedure in Algorithm \ref{alg: boot_ERGM}.

\begin{algorithm}
\SetAlgoLined
\KwIn{Observed sociomatrix: $G$; Bootstrap iterates: $B$.}
Compute an estimate of $\theta$, denoted by $\hat \theta$\;

\For{$b = 1$ \KwTo $B$}{
Sample $G_{k}^* \sim F_{\hat \theta}$;

Set $A_k^\star$ to be the $n \times n$ adjacency matrix for the graph $G^\star_k$\;

Record $A^*_k$\;
}
For all $i, j$, compute $$\hat P_{ij} = \frac{1}{B}\sum_{k = 1}^B [A^*_{k}]_{ij} \ ;$$

\KwOut{Estimated probability matrix: $\hat{P}$.}

 \caption{Given sociomatrix $G$, simulate $\hat P$}
 \label{alg: boot_ERGM}

\end{algorithm}

Having now described how to estimate $P$ from an estimate of the ERGM parameter, we now consider an ERGM model and show how to test the significance of its parameters. Consider the model 
\begin{equation}
\label{eq: model_ERGM_test}
    P(G = g) \propto \exp\left(\theta_1 \cdot \text{edges} + \theta_2 \cdot \text{triangle} + \theta_3 \cdot \text{kstar(2)} \right) \;,
\end{equation}
\label{eq: ergm_model}
where $\text{edges}$ counts the number of edges in $g$, triangles counts the number of triangles, and k-star(2) counts the number of 2-stars, which is a triangle with one edge missing.

\begin{figure}
    \centering
    \includegraphics[scale = 0.5]{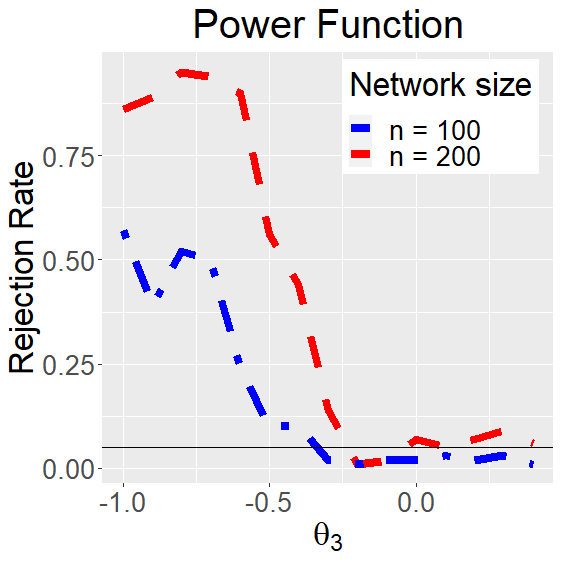}
    \caption{\footnotesize{Power function for the hypothesis in (\ref{eq: null_ERGM}). The null hypothesis is $\theta_3 = 0$. The black horizontal line represent the $\alpha = 0.05$ threshold.}}
    \label{fig: power_ERGM}
\end{figure}

Suppose that we are interested in testing whether $\theta_3 = 0$. In other words, we believe that the model above is correctly specified, with the exception that we do not know if $\theta_3 \neq 0$. Writing this as a hypothesis testing problem, we want to test the hypothesis
\begin{equation}
    H_0: \theta_3 = 0,  \ \ \ H_a: \theta_3 \neq 0 \;.
    \label{eq: null_ERGM}
\end{equation}
To test this, we fit our data to the model in (\ref{eq: model_ERGM_test}) with $\theta_3 = 0$. That is, we estimate $(\theta_1, \theta_2)$ in the model $P(G = g) \propto \exp\left(\theta_1 \cdot \text{edges} +\theta_2 \cdot \text{triangle}\right)$. Let $(\hat \theta_1, \hat \theta_2)$ denote these estimates. We then simulate $\hat P$ using Algorithm \ref{alg: boot_ERGM}. With these estimates, we can then form the matrix $\hat A = (G - \hat P)/\sqrt{(n-1)\hat P(1-\hat P)}$, with $\hat A_{ii} = 0.$ We test $H_0$ using Algorithm \ref{alg: boot_undirected}. In Figure \ref{fig: power_ERGM}, we plot the power function for the hypothesis. We see that near $\theta_3 = 0$, the power is roughly equal to the Type 1 error $\alpha = 0.05$. As $|\theta_3|$ becomes larger, the power increases. We also see that the power increases for all $\theta_3 \neq 0$ as $n$ increases.

\subsection{Testing Degree Heterogeneity Using ARD}
\label{sec: ARD}

In this section, we show an interesting application of our method to the case of partial network data. We focus on a particular type of partial network data known as Aggregated Relational Data (ARD). This type of data is often cheaper to collect and can still be used to perform inference. For example, \cite{Breza_consist} showed that the maximum likelihood estimate (MLE) for the latent space model, computed using only ARD instead of the entire network, is consistent as the graph size grows.

Let $G$ denote a network of interest on $n$ nodes and suppose that we want to test if there is degree heterogeneity in the network. One way to model this question is through the following:
\begin{equation}
    H_0: g \sim \text{ER}(p^\star) \text{ for some } p^\star, \ \ \ H_a: \  H_0 \text{ is false} .
\end{equation}
where ER$(p^\star)$ denotes an Erdös-Rényi model with unknown parameter $p^\star.$ Suppose that instead of observing the whole network $g$, we instead observe Aggregated Relational Data (ARD). 

Under the null hypothesis, each $Y_{ij} \sim \text{Binomial}(n_j, p^\star)$, where $n_j = |G_j|$ is the size of group $G_j$. So if we define an $m \times K$ matrix $A$ with 
\begin{equation*}
    A_{ij} = \frac{Y_{ij} - n_j p^\star}{\sqrt{n_j p^\star(1-p^\star)}} \;,
\end{equation*}
then $A$ is a $m \times K$ random matrix with mean zero and variance 1. Note that unlike in previous forms of $A$, in this case the diagonal of $A$ is not set to be zero.

In general, we do not know $p^\star$ but given ARD, we can estimate $p^\star$  with
\begin{equation*}
    \hat p = \frac{1}{mK} \sum_{i = 1}^{m} \sum_{j = 1}^{K} \frac{Y_{ij}}{n_j} \;.
\end{equation*}
Under $H_0$, since $E(Y_{ij}/n_j) = p^\star$, it follows that $\hat p \overset{p}{\rightarrow} p^\star$ as $m \rightarrow \infty$. Here we consider $K$ fixed; see the discussion at the end of Section \ref{sec: partial}. We can therefore define
\begin{equation*}
    \hat A_{ij} = \frac{Y_{ij} - n_j \hat p}{\sqrt{n_j \hat p(1-\hat p)}} \;,
\end{equation*}
We can use Theorem \ref{thm: asymm} to construct a test statistic for the null hypothesis. Our test statistic is the largest singular value of the matrix $\hat A$. Our rejection region for the null hypothesis is based on the quantiles of the Tracy-Widom distribution, as indicated in Theorem \ref{thm: asymm}.

\begin{figure}
    \centering
    \subfloat[\centering ]{{\includegraphics[width=7cm]{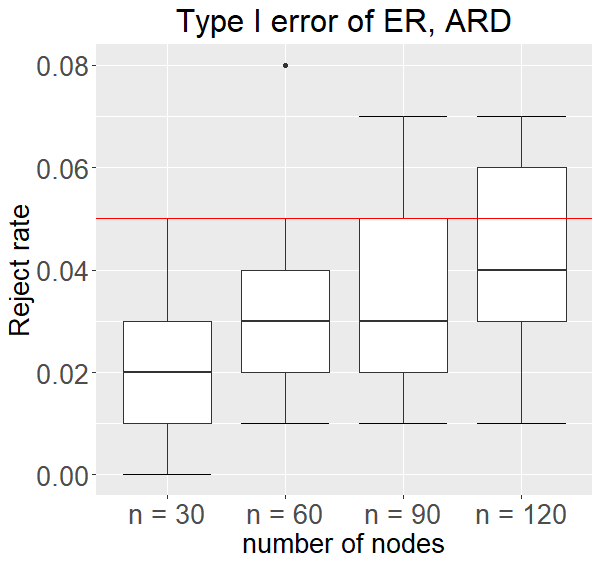} }}%
    \qquad
    \subfloat[\centering]{{\includegraphics[width=7cm]{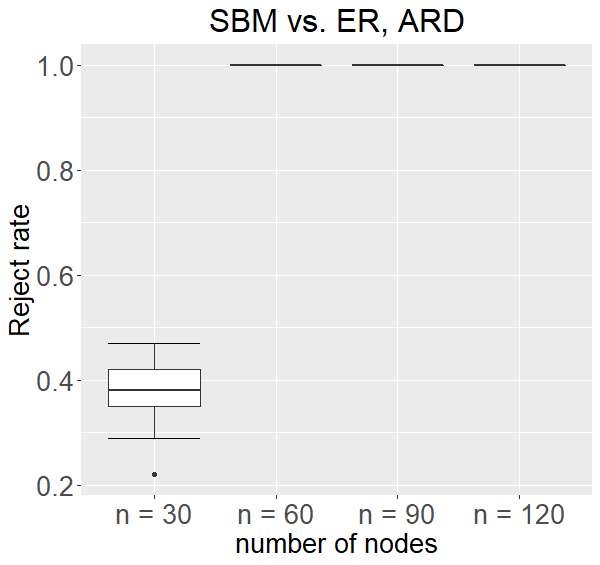} }}%
    \caption{\footnotesize{Left: Type I error of ER model via ARD. Right: Power of fitting SBM ARD to ER model. When the hypothesis model is correct, we observed a Type I error centered around the level of testing $\alpha = 0.05$. When the ARD of a more complex model is fitted to a simple hypothesis model (i.e. ER is a special case of SBM with one community), we will observe a very high power which grows with network size $n$. }}%
    \label{fig: ARD}%
\end{figure}

We first consider the Type I error of this method. For $n \in \{30, 60, 90, 120\}$, we draw an Erdös-Rényi graph with $m = \gamma_m n$ and $K = \gamma_K n$, where $\gamma_m = 1/3$ and $\gamma_K = 1/10$. We divide nodes equally into each of the $K$ categories. Given a graph $G$, we define $Y_{ij}$ as in (\ref{eq: ARD_def}). Our goal is to test whether $G$ is drawn from an ER model. 
We plot our results in Figure \ref{fig: ARD}.

Of course, more complicated testing problems can be used, but we leave that to future work. The goal of this section is to simply show how our method might be used to analyze network goodness-of-fit in cases where only partial network data is available.

\subsection{Directed Network Case}
\label{sec: directed}

In this section, we show how to test (\ref{eq: origina_H}) when the network is directed. Recall that Theorem \ref{thm: asymm} and \ref{thm: exp_directed} tell us the distribution of singular values and so they provide us with test statistics. 

Suppose that we are given a directed graph $g$. We are interested in testing whether $g$ is drawn from a directed Erdös-Rényi model. By this, we mean a directed graph whose directed edges form independently with probability $p^\star$. Our goal is to test
\begin{equation}
    H_0: G \sim \text{DER}(p^\star) \text{ for some } p^\star, \ \ \ H_a: H_0 \text{ is false.} 
    \label{eq: H_0_Dir}
\end{equation}
where the notation $\text{DER}(p)$ stands for a directed ER model. Theorems \ref{thm: asymm} and \ref{thm: exp_directed} give us test statistics to test this hypothesis. We start with the statistic from Theorem \ref{thm: asymm}. This theorem states, informally, that the singular values of $X$, once rescale and re-centered, converge to a Tracy Widom distribution. As in the undirected case, this convergence can be slow, so we use the bootstrap correction algorithm in Algorithm \ref{alg: boot_dir}.  Theorem \ref{thm: exp_directed} also provides a test statistic to test (\ref{eq: H_0_Dir}). This theorem states, informally, that $n$ times the largest singular value of a random matrix converges to an exponential random variable. 

\begin{figure}
   \centering
   \subfloat[][]{\includegraphics[width=.33\textwidth]{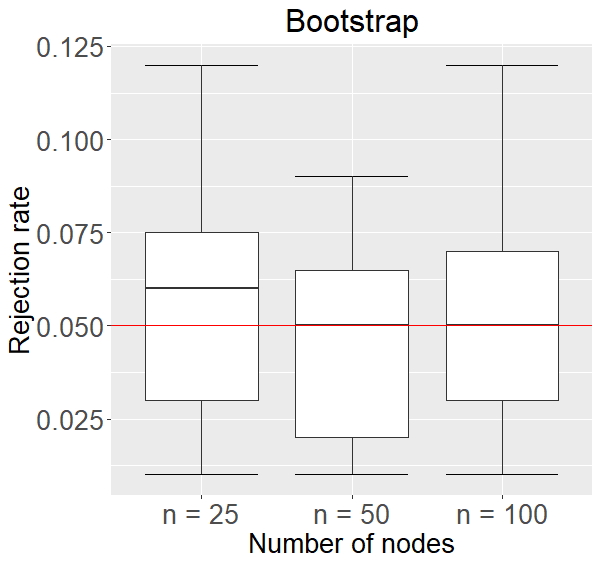}} 
   \subfloat[][]{\includegraphics[width=.33\textwidth]{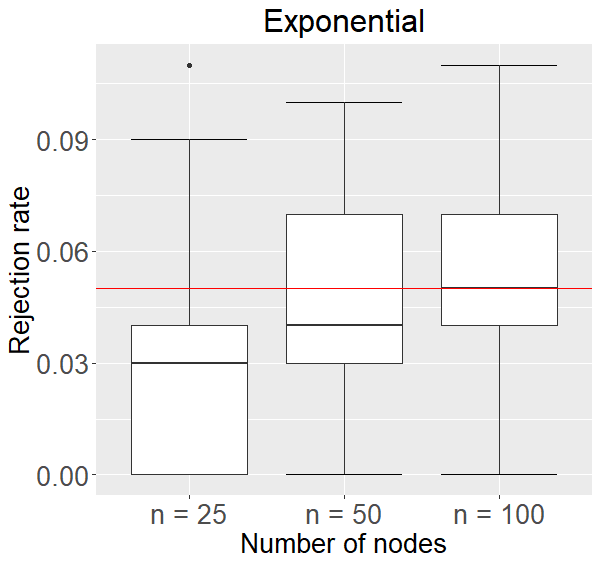}} 
   \subfloat[][]{\includegraphics[width=.33\textwidth]{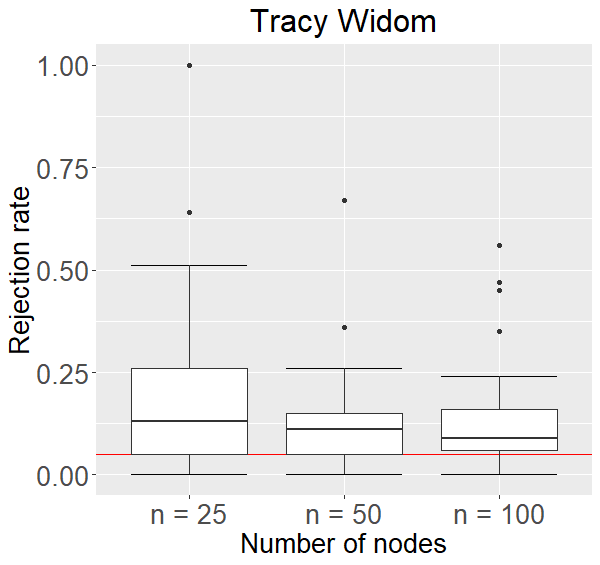}} 
   \quad
   \subfloat[][]{\includegraphics[width=.33\textwidth]{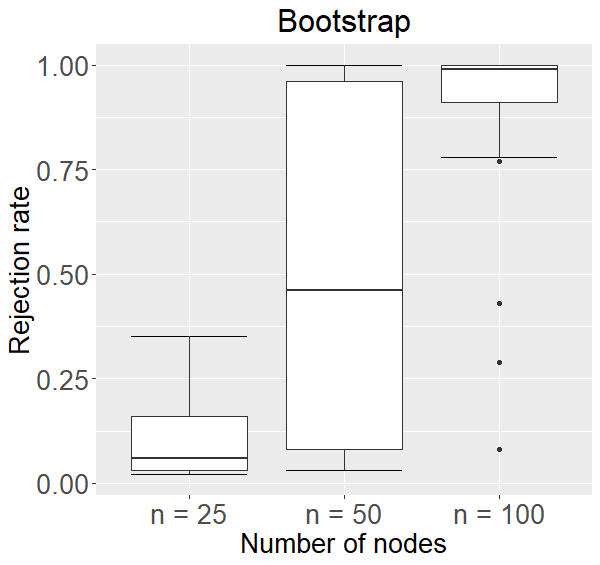}}
   \subfloat[][]{\includegraphics[width=.33\textwidth]{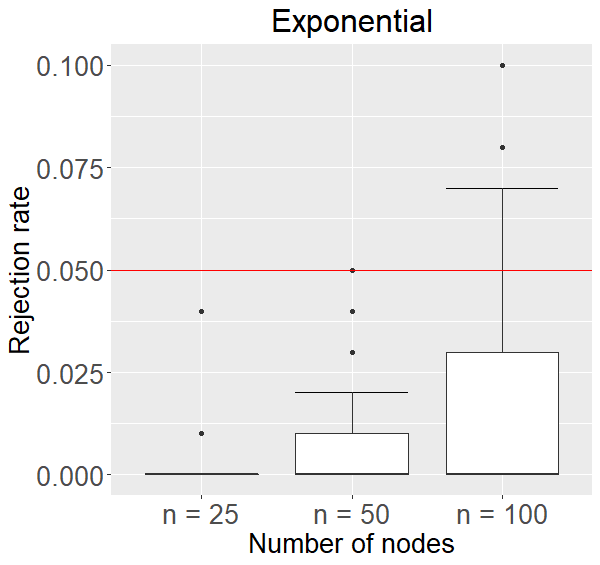}}
   \subfloat[][]{\includegraphics[width=.33\textwidth]{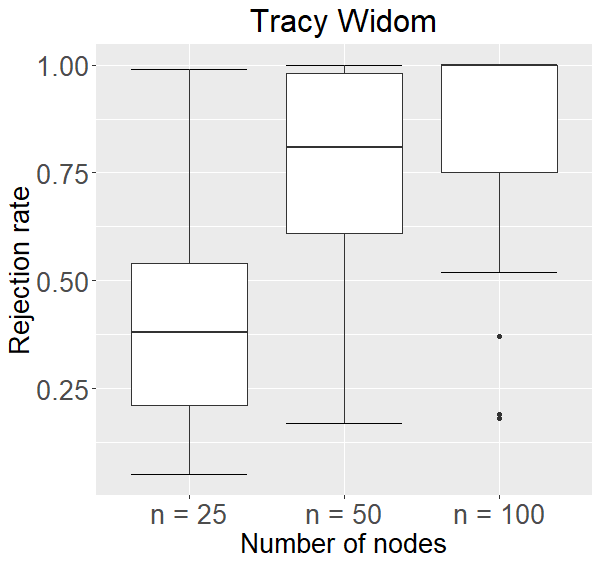}}
   \caption{\footnotesize{Type 1 error and rejection rates for directed network data. The first row corresponds to the case of a directed Erdös-Rényi model. In the top left figure, we plot the average rejection rate over 50 sets of simulations for $n = 25, 50, 100$ using the bootstrap test from Section \ref{sec: Boot_Directed} . In the top middle, we plot the average rejection rate using the exponential test statistic in Theorem \ref{thm: exp_directed}. In the top right, we plot the average rejection rate using Tracy Widom test statistic in Theorem \ref{thm: asymm}.  In the second row, we plot the average rejection rates using a directed stochastic block model (DSBM) with  2 communities and distinct cross community probabilities. We see that bootstrap and exponential methods have good Type I error, yet that of Tracy Widom statistics are relatively larger. In terms of power against DSBM, bootstrap and Tracy Widom obtain good power, but the Exponential does not. Overall, the bootstrap statistic has a better performance in general.}}
   \label{fig: directed}
\end{figure}

Using these two theorems, we can test $H_0$ in (\ref{eq: H_0_Dir}). In Figure \ref{fig: directed}, we plot the type 1 Type I error in the first row for the ``bootstrap" method from Algorithm \ref{alg: boot_dir} and the ``exponential" method. The second row plots the power of our method when $g$ is drawn from a directed stochastic block model with two communities. We see that both methods have a good control on the Type I error at $\alpha = 0.05$, but only the ``bootstrap" method is able to distinguish between a DER and a directed stochastic block model.

\section{Community Detection with latent space models}
\label{sec: real}

In this section, we analyze three data sets that are studied in \cite{Ma2020}: the Political Blog data, the Simmons College data, and the Caltech data. \cite{Ma2020} fits these data sets to the latent space model in (\ref{eq: LS_model}) without covariate components.

The authors computed the estimates of the latent space positions $\{z\}$ with the projected gradient descent methods, then applied a simple $k$-means clustering on the estimated latent positions for community detection. In Table 1 of their work, \cite{Ma2020} compares the clustering results with the community membership provided in the original data set, and reported the mis-classification rate between the estimated clustering and the true network clustering. In this analysis, the fitted latent dimensions are set to either $K$ or $K + 1$, where $K$ is the known number of clusters in the data. We observed that fitting these datasets to different dimensions changed the mis-classification rate, which suggests that choosing an optimal latent dimension is crucial for community detection.

We made three major adjustments based on their evaluating procedure. First, instead of directly setting the latent dimension as $K$ or $K + 1$, we fit the data sets with Algorithm \ref{alg: dim_pred} and used the resulting $d_{\text{fit}}$ as the fitted dimension. Second,  the $k$-means method produces different clustering results even with the \texttt{replicate} command, so to avoid bias, we run the $k$-means clustering function 200 times in MATLAB and select the set of positions with the best fit. We then repeat this process 100 times and return the average mis-classification rate across the 100 simulations. Lastly, instead of using only the first $k$ eigenvectors of $\hat{z}$ as in \cite{Ma2020}, we simply use the estimated positions $\hat z$ in the $k$-means algorithm. Our approach is intuitive, simple, and yields good performance on these three data sets. 

We present our results in Table \ref{tab: real_data} and Figure \ref{fig: mis_cla}.  For Table \ref{tab: real_data}, in column $t_{\text{TW}}$'s, text labelled with star indicates the Tracy Widom test statistics is not rejected. In column $R_{\text{mis}}$, bold text indicates the optimal classification rate. Figure \ref{fig: mis_cla} gives a visual representation of the misclassification rate over different choices of latent dimensions.

In the Political Blog data and Caltech data, the optimal dimension chosen by our method are 7 and 8 respectively. The test statistics for $d_{\text{fit}} > d_{\text{opt}}$ are also not rejected. This behavior is similar to the behavior we saw in the latent space simulations. The optimal mis-classification rates are also achieved at $d_{\text{opt}}$. Compared to the results in Table 1 of \cite{Ma2020}, for the Political Blog data, we obtained a better mis-classification rate, from $4.513 \%$ (latentnet) to $4.26\%$ (Latent Space based Community Detection (LSCD), $d_\text{fit} = 7$). For the Caltech data, we obtained the same optimal rate $18.35 \%$ (LSCD, $d_\text{fit} = 8$), as our predicted dimension coincides with the number of clusters.  These results shows that the latent space model is a good fit of the two data sets, and our method performs well in achieving the optimal mis-classification rate.

In the Simmons College data, the predicted dimension is $d_{\text{opt}} = 8$, with mis-classification rate $10.37 \%$. However, for $d_{\text{fit}} > d_{\text{opt}}$, we still observe that some fitted dimensions, namely $d = 10, 12$, are rejected. Moreover, the result for the Tracy Widom statistics is not as robust as in previous two cases: our algorithm provides different predicted dimensions in different trials, whereas the results are consistent in the previous two data sets. This potentially suggests that the latent space model might not be a good fit for the Simmons College data. Nevertheless, our method still reveals certain natures of the network. The optimal rate is achieved at $d_{\text{fit}} = 11$, which is also substantially larger than the fitted dimension $d_{\text{fit}} = 4$ in \cite{Ma2020}, at which our test statistic is not rejected. The mis-classification rate is improved from $11.17 \%$ (LSCD, $d_\text{fit} = K + 1$) to $9.62 \%$ (LSCD, $d_\text{fit} = 11$).

Our result shows that, based on the behavior of the test statistics with $d_{\text{fit}} > d_{\text{opt}}$, our Algorithm \ref{alg: dim_pred} potentially suggests whether the latent space model can be a good fit for the observed network. For networks that fit the latent space model well, our method will choose the optimal latent dimension that minimizes the community detection misclassification rate.

\section{Conclusion}
In this work we proposed a network goodness-of-fit test that uses the eigenvalues of the centered, scaled adjacency matrix. We used recent work in random matrix theory to derive a test statistic that can test whether an observed network is a good fit for common network models. This framework can handle undirected and directed networks, and can also handle cases where the researcher only has access to partial network data. We discussed the performance of this method on several common network models, like the latent space model, and showed that the test has favorable properties in terms of Type 1 error and power. 

There are many avenues of future work. First, we would like to answer more general goodness-of-fit questions when the researcher only has access to ARD. We believe that the estimation methods presented in \cite{Leung} can be used to estimate the $m \times K$ matrix $P$ under a variety of realistic null hypotheses, which means that we can test the null hypothesis in (\ref{eq: origina_H}) in a variety of more realistic settings.  Second, we would like to extend this method to time-varying networks, such as those considered in \cite{Townsend}. Finally, we would like to determine whether other random matrix theory results, such as Theorem 1 in \cite{Komlos}, will lead to a test of (\ref{eq: origina_H}) with better properties, like higher power.

\bibliographystyle{apalike}
\bibliography{References, Second_LS_References, References_LSPaper}

\pagebreak
~\\~\\
\centerline{\huge Appendix}~\\
\appendix 
\section{Bootstrap Correction for Directed Data}
\label{sec: Boot_Directed}
\renewcommand\thefigure{\thesection.\arabic{figure}}  
\setcounter{figure}{0} 
\renewcommand\thetable{\thesection.\arabic{table}}  
\setcounter{table}{0}
We now give a bootstrap correction algorithm for directed network data.

\begin{algorithm}[H]
\SetAlgoLined
\KwIn{Observed sociomatrix: $G$; Estimated probability matrix $\hat{P}$; Bootstrap iterates: $B$; $TW_1$ mean: $\mu_{TW}$; $TW_1$ standard deviation: $s_{TW}$; Significance Level: $\alpha$.}
 Compute 
    \begin{equation*}
        \hat A_{ij} = (G_{ij} - \hat P_{ij})/\sqrt{\hat P_{ij}(1 - \hat P_{ij})}\ ;
    \end{equation*}
\For{$b = 1$ \KwTo $B$}{
Sample $G^\star_b \sim F_{\hat \theta}$\;

Compute
    \begin{equation*}
         [A^\star_b]_{ij} = ([G^\star_b]_{ij} - \hat P_{ij})/\sqrt{\hat P_{ij}(1 - \hat P_{ij})}\ ;
    \end{equation*}
    
Set $\lambda_b^\star = s_{\max}(A^\star_b)$\;
}

Define $ \mu$ to be the sample mean of the $\{\lambda_b^\star\}_{b = 1}^B$ and $s$ to be the sample standard deviation of $\{\lambda_b^\star\}_{b = 1}^B$\;

Compute the test statistic $t$
\begin{equation*}
    t := \mu_{TW} + s_{TW} \left(\frac{s_{\max}(\hat A)- \mu}{s}\right).
\end{equation*}

\uIf {$TW_1(\alpha / 2) < t < TW_1(1 - \alpha / 2)$} {
    Do not reject $t$ and set $\text{Rej} = \text{FALSE}$\;
} \Else {
    Reject $t$ and set $\text{Rej} = \text{TRUE}$.
}

\KwOut{Rejection of bootstrap statistic: Rej.}
 \caption{Bootstrap correction of Directed Tracy Widom statistic}
 \label{alg: boot_dir}
\end{algorithm}

\section{Using BIC to select dimension of latent space}
\label{sec: LS_Appendix}
Suppose $G$ is a network with $n = 100$ nodes drawn from a latent space model, defined in \citet{hoff2002latent}, where its latent space dimension is $d_{\text{true}} = 2$. We fit the observed network $G$ to the latent space models with dimensions $d_{\text{fit}} = {1, 2, 3, 4}$ and calculate the corresponding BIC with the \texttt{latentnet::ergmm.bic} command. We summarize the results in Table \ref{table: 1}. The BIC method provides a false prediction, suggesting the dimension to be $d_{\text{fit}} = 4$ instead of $d_{\text{true}} = 2$. This indicates the BIC method might not be an optimal approach for latent space dimension detection as stated in the \texttt{latentnet} manual: ``\textit{It is not clear whether it is appropriate to use this BIC to select the dimension of latent space ...}"
This motivates us to develop Algorithm \ref{alg: dim_pred} in Section \ref{sec: LS} to robustly address the problem. Our method correctly predicts the latent space dimension 80$\%$ of the time or better for a variety of true dimensions and for values of $n$ (the number of nodes) as small as 100. Crucially, this sample size covers many empirically-relevant networks, such as the Indian villages network studied in \cite{breza2017using} and others. 
\\
\begin{table}[H]
\centering
\renewcommand{\arraystretch}{1.2}
 \begin{tabular}{c c c c c} 
 \hline
 & $d_{\text{fit}} = 1$ & $d_{\text{fit}} = 2$ & $d_{\text{fit}} = 3 $& $d_{\text{fit}} = 4$ \\
 \hline
 BIC&6047.10& 5774.94 & 5750.63& {\color{red} 5721.85} \\
 \hline
 \end{tabular}
 \vspace{10pt}
 \caption{\footnotesize{Fitted BIC of the observed network $G$ with $d_\text{true} = 2$, which suggest $d_{\text{fit}} = 4$ be the underlying latent dimension. }}
\label{table: 1}
\end{table}

\begin{table}[ht!]
\centering
\setlength{\tabcolsep}{10pt} 
\renewcommand{\arraystretch}{1.5}
\begin{tabular}{ccccccc}
 \hline
                 & \multicolumn{2}{c}{Political Blog} & \multicolumn{2}{c}{Simmons College} & \multicolumn{2}{c}{Caltech}    \\
                 \hline 
$d_\text{fit}$ & $t_{\text{TW}}$ & $R_{\text{mis}}$ & $t_{\text{TW}}$ & $R_{\text{mis}}$   & $t_{\text{TW}}$ & $R_{\text{mis}}$\\
                1 &        2.23 &         5.16       &      48.03    &          16.00        &    7.94  &           56.17     \\
                2 & 2.70        &   4.58             &   32.12       &          15.39           &   8.31  &     33.41          \\
                3 &     12.20    &      4.34         &      17.95    &          18.21           &   13.82&      37.37           \\
                4 &    15.11     &      4.42            &  19.99       &        14.89           &   7.15&  36.57        \\
                5 &     10.80    &          4.91        &      16.18   &        11.13            &  5.79 &  31.48     \\
                6 & 8.67        &                4.58   &     8.29     &        10.30             & 2.90  &     21.11       \\
                7 &     $-1.75^*$    &  \textbf{4.26}        &      2.29   &        10.28             & 2.47  &         27.57      \\
                8 &     $-2.78^*$    &                  4.42    &   $1.14^*$   &        10.37             & $-0.83^*$ &\textbf{18.35}          \\
                9 &     $-1.33^*$    &                      5.07&   $0.84^*$   &        9.87             &  $0.06^*$ &         19.27     \\
                10 &    $-1.76^*$     &                5.32     &   2.37   &              9.67     & $-0.73^*$&  19.40         \\
                11 &        $-1.26^*$ &                5.33     &   $0.30^*$   &    \textbf{9.62}         &$0.46^*$ &  19.07         \\
                12 &    -   &           -                 &     1.92 &               10.21          &  -&       -    \\
                \hline
\end{tabular}
\vspace{5pt}
\caption{\footnotesize{Tracy Widom statistics and mis-classification rates of Political Blog data, Simmons College data, and Caltech data. Tracy Widom statistics that are not rejected are labelled with stars. Optimal mis-classification rates are highlighted in bold text. }}
\label{tab: real_data}
\end{table}

\begin{figure}[ht!]
\centering
\subfloat[Political Blog data]{\includegraphics[width=.33\linewidth]{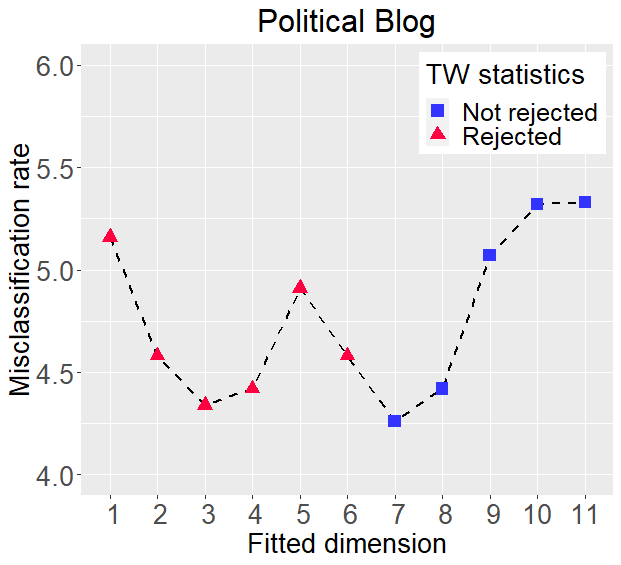}}
\subfloat[Simmons College data]{\includegraphics[width=.33\linewidth]{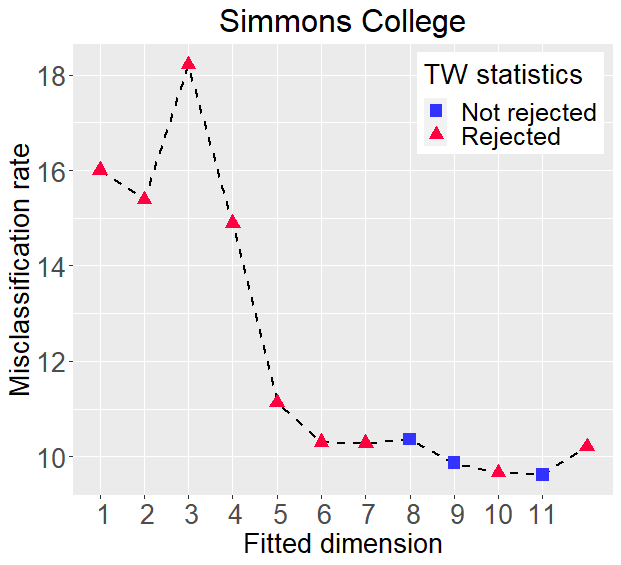}}
\subfloat[Caltech data]{\includegraphics[width=.33\linewidth]{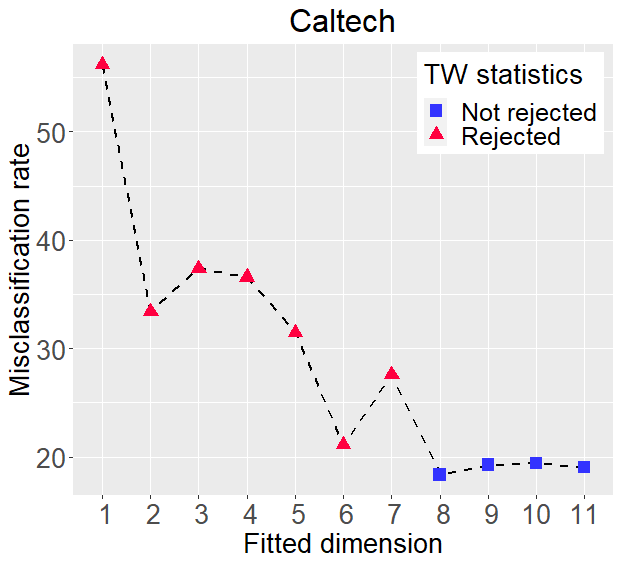}}
\caption{\footnotesize{Mis-classification rates of Political Blog data, Simmons College data, and Caltech data.}}
\label{fig: mis_cla}
\end{figure}

\begin{figure}
    \centering
    \subfloat[\centering ]{{\includegraphics[width=7cm]{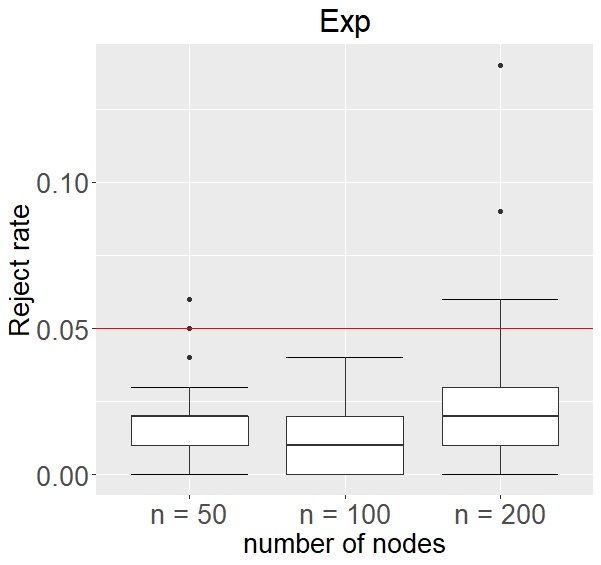} }}%
    \qquad
    \subfloat[\centering]{{\includegraphics[width=7cm]{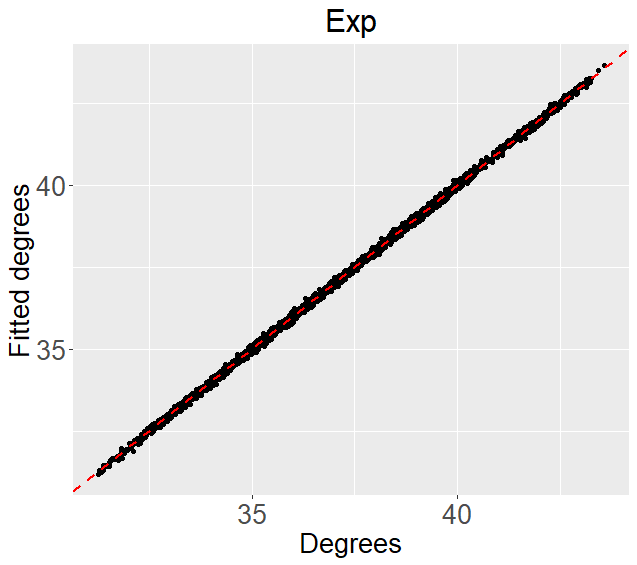} }}%
    \caption{\footnotesize{Left: Power for the null hypothesis in (\ref{eq: beta_null}) against Beta model with exp link function for $n = 50, 100, 200$. The powers centered below $0.05$ and is smaller than the corresponding Type I error. Right: Identical settings as in Figure \ref{fig: expit_data}, with true model altered to exp link function. Surprisingly, we observed that most of the points align upon the diagonal, which potentially indicates that the exp model can also be a good fit. Such a phenomenon is observed with other network statistics, i.e. average path length, number of 3/4-cliques, etc, which suggests there might exist an equivalent relationship between the expit and exp link function.}}%
    \label{fig: exp_data}%
\end{figure}

\begin{figure}
    \centering
    \subfloat[\centering ]{{\includegraphics[width=7cm]{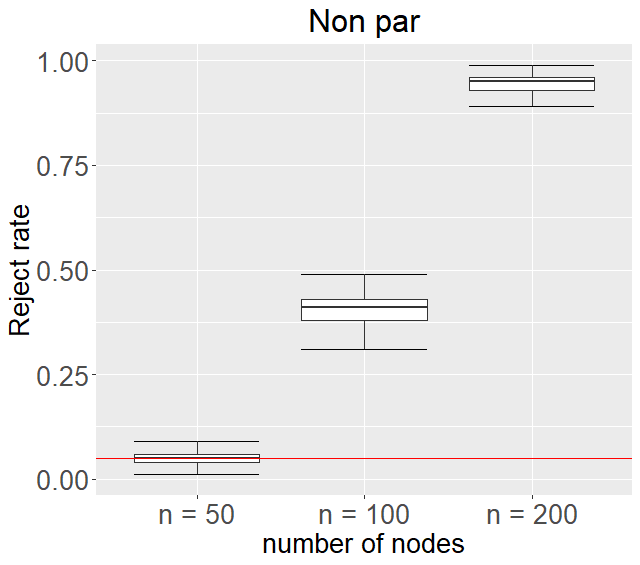} }}%
    \qquad
    \subfloat[\centering]{{\includegraphics[width=7cm]{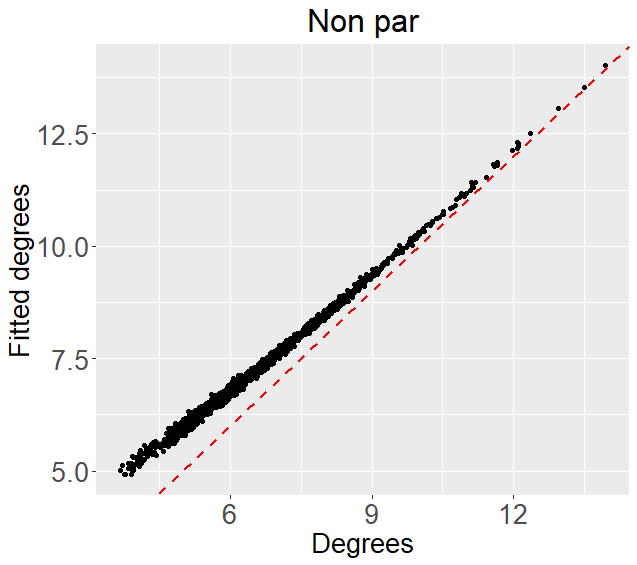} }}%
    \caption{\footnotesize{Left: Power for the null hypothesis in (\ref{eq: beta_null}) against non-parametric network structures for $n = 50, 100, 200$. The powers increases sharply as network sizes grows. Right: Identical settings as in Figure \ref{fig: expit_data}, with true model altered to non-parametric structures. We observed that the trend of the points tilts up at the left end, with more mass concentrates around smaller degree distributions. Such a behavior differs significantly with that of the $\beta$-model with expit link function, which is consistent with our observation on the left that our method reject almost 100\% of the time for $n = 200$.}}%
    \label{fig: non_par_data}%
\end{figure}

\begin{figure}
\centering
\subfloat[Beta model with expit link function]{\includegraphics[width=.45\linewidth]{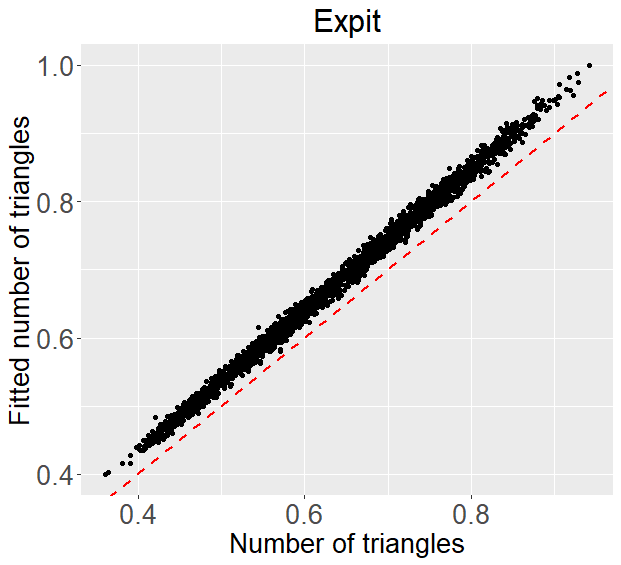}} \hfill
\subfloat[Beta model with exp link function]{\includegraphics[width=.45\linewidth]{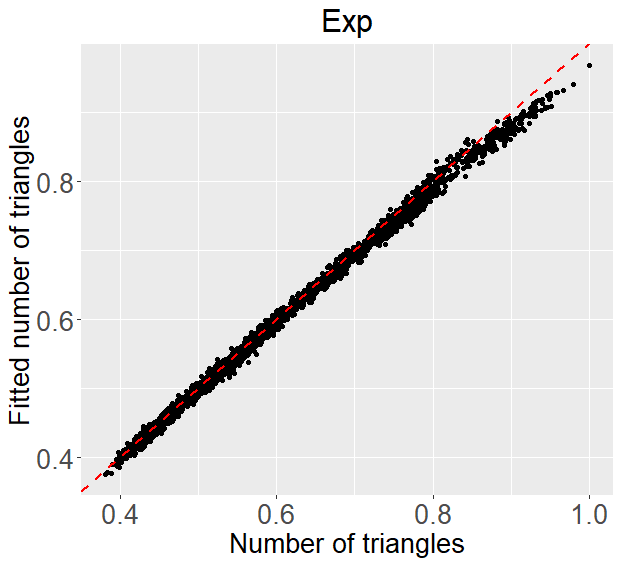}} \par 
\subfloat[Non-parametric structure ]{\includegraphics[width=.45\linewidth]{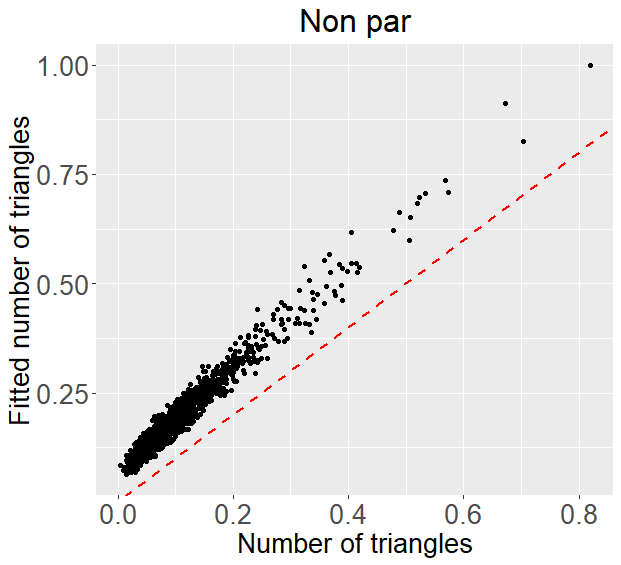}}
\caption{\footnotesize{We plot the number of triangles in observed networks against the number of triangles simulated via fitted MLE estimates w.r.t. expit link function. The red dash line corresponds to $y  =x$. If the fit is good, we will observe the data points align upon $y = x$. Compared to the poorly behaved non-parametric structure, we observed a good correspondence between the observed and fitted on Beta model with expit and exp function, indicating the goodness-of-fit of the two models is probably good. This further tell us there is potentially an equivalent relationship between the two link functions and can be achieved with the fixed point method in \cite{diaconis}. The difference between the simulated values in black and the diagonal line in (A) decreases as the sample size $n$ increases.}}
\label{fig: num_triangles_beta}
\end{figure}

\end{document}